\documentclass[aps,pra,twocolumn,amsfonts,amsmath,amssymb,superscriptaddress]{revtex4-1}
\usepackage{graphicx}\usepackage[ colorlinks = true,
             linkcolor = blue,
             urlcolor  = blue,
             citecolor = red,
             anchorcolor = green,
]{hyperref}\usepackage{graphicx}
\usepackage{bbold}

\newcommand*{\vv}[1]{\vec{\mkern0mu#1}}

\begin{document}

\title{Franck-Condon factors via compressive sensing}

\author{Kevin Valson Jacob} 
\author{Eneet Kaur}
\address{Hearne Institute for Theoretical Physics and Department of Physics and Astronomy,
Louisiana State University, Baton Rouge, Louisiana 70803, USA
}
\address{
National Institute of Information and Communications Technology, 
Koganei, Tokyo 184-8795, Japan
}
\author{Wojciech Roga}
\author{Masahiro Takeoka}
\address{
National Institute of Information and Communications Technology, 
Koganei, Tokyo 184-8795, Japan
}

\date{\today}

\begin{abstract}

Probabilities of vibronic transitions in molecules are referred to as Franck-Condon factors (FCFs). Although several approaches for calculating FCFs have been developed, such calculations are still challenging. Recently it was shown that there exists a correspondence between the problem of calculating FCFs and boson sampling. However, if the output photon number distribution of boson sampling is  sparse then it can be classically simulated. Exploiting these results, we develop a method to approximately reconstruct the distribution of FCFs of certain molecules. We demonstrate this method by applying it to formic acid and thymine at $0$ K.
In our method, we first obtain the marginal photon number distributions for pairs of modes of a Gaussian state associated with the molecular transition. 
We then apply a compressive sensing method called polynomial time matching pursuit to recover FCFs. 
\end{abstract}

\maketitle

\section{Introduction}

In theoretical molecular spectroscopy, the goal is to obtain better understanding of changes in molecular structure and force field due to transitions by analyzing  spectra of molecules. Of importance is the vibronic spectra of molecules, where the spectral lines correspond to transitions between two vibrational levels of electronic states. The probabilities of transitions between these levels is given by Franck-Condon factors (FCFs) which determine  the intensities of the spectral lines. Given a vibronic spectrum, knowing which transitions contribute to observed spectral lines is helpful in testing a variety of molecules before synthesizing the most appropriate one for a particular application. Such applications include synthesizing new drugs and increasing the efficiency of solar cells among others \cite{hachmann,butler,gross2000}. Therefore, it is relevant to develop algorithms to compute FCFs.

According to the Franck-Condon principle, the electronic levels of molecules are modelled as multi-mode quantum harmonic oscillators (HO). Therefore, the vibronic wavefunctions are multi-dimensional Hermite polynomials; and their overlaps are the transition probabilities (FCFs). There are infinitely many possible transitions from a given ground state to excited states. Moreover, even if we truncate the maximum number of excited states, the number of transitions scales exponentially with the number of atoms of the molecule. Also, calculating the overlap of Hermite polynomials is computationally challenging \cite{philips2019}. These reasons contribute to the difficulty in simulating the vibronic spectra of molecules. 

There exists a classical recursive method to generate FCFs developed by Doktorov \cite{Doktorov}. However, this method is inefficient with respect to both computing time and memory required. Recently, several other approaches \cite{Rabidoux,Jankowiak,santoro,Ruhoff} have been developed which improve upon previous methods. Another work \cite{quesada2018} explores the connection between FCFs and graph theory.

\par
In this paper, by applying compressive sensing \cite{Candes2,Donoho,Candes1,Baraniuk,Candes3}, we develop an alternate method to approximately reconstruct the distribution of FCFs. Compressive sensing techniques, introduced in the seminal papers \cite{Candes2,Donoho,Candes1,Baraniuk,Candes3}, are particularly useful for recovering sparse distributions from compressively sensed data. These techniques provide an efficient procedure to reconstruct a large but sparse data set. In such techniques, compressively sensed data $y$ is linearly related to a large unknown data set $x$ via a measurement matrix $A$, i.e. $y=Ax$. Although this system of linear equations is underdetermined, owing to the sparsity of the dataset to be reconstructed, various algorithms \cite{mallat,davis,Candes2,gradientpursuit,blumensath2,Draganic,zhang} for efficient recovery of $x$ have been developed. These techniques have then been applied in various fields such as image reconstruction \cite{romberg}, communications \cite{mihajlovic}, medical imaging \cite{lustig}, and quantum states tomography \cite{gross}. 

In \cite{guzik2015}, following the work of \cite{Doktorov}, it was noted that there exists a parallel between molecular vibronic transitions and Gaussian boson sampling \cite{Hamilton2017}. In this paradigm, it is possible to express the excited state as a Gaussian state \cite{weedbrook}, which is obtained by a Gaussian evolution of the ground state. FCFs can then be obtained as the probabilities of various photon number configurations of this excited state. However, finding the probabilities of various photon number distributions of a Gaussian state is known to be in the complexity class $\#\mathrm{P}$ \cite{quesada2018,Kruse}.

If the photon number distribution of the Gaussian state is sparse, then it is possible to recover this distribution approximately. This is possible by calculating the marginal photon number distributions over a few modes of the Gaussian state and then applying compressive sensing techniques to recover the full distribution. This idea was first explored in \cite{roga}. The calculation of marginal photon number distributions over a few modes of the Gaussian states is computationally tractable. By employing a technique developed in \cite{quesada2018}, the probabilities of marginal photon number distributions can be calculated as loop-Hafnians of certain matrices related to the Gaussian state (Section~\ref{sec:photonnumber}).

Our contribution is to develop a method to approximately reconstruct the distribution of FCFs. To this end, we apply \textbf{polynomial time matching pursuit} (PTMP) \cite{roga} defined in Section~\ref{sec:ptmp}. This is a first-order greedy algorithm which is a modification of the matching pursuit algorithm \cite{mallat}. With this modification, the runtime of PTMP is polynomial in the number of modes of the excited Gaussian state. Since the number of modes is in general $3N_a-6$ where $N_a$ is the number of atoms of the molecule, PTMP is efficient when dealing with large molecules. 

The key step of PTMP, which is also its computational bottleneck, is the identification of a column of the measurement matrix $A$ which has the largest overlap with a specified vector in each iteration of the algorithm. This maximization procedure would typically take the time of the order of the size of the matrix $A$. However, if only the marginal photon number distributions of nearest neighbor modes are considered, then the problem can be mapped to an optimization in a 1-D Ising model \cite{roga}. The task is then to find the highest energy configuration of a 1-D classical spin chain with the Ising Hamiltonian which can be efficiently obtained.

In Section~\ref{sec:results}, we demonstrate the feasibility of PTMP and analyze its performance in practice. We do this by approximately reconstructing the distributions of FCFs of two molecules: formic acid (7 mode symmetry block) in Section~\ref{sec:formicacid} and thymine (26 mode symmetry block) in Section~\ref{sec:thymine}. 

\section{Methods}

\subsection{Gaussian transformation corresponding to a vibronic transition}\label{sec:gaussian}

Let us introduce a general formalism for describing bosonic Gaussian states  \cite{Braunstein,Eisert,Braunstein2,Ferraro}. We write the boson creation and annihilation operators for a given mode $u$ as  $\hat{a}_u^\dagger$ and $\hat a_u$ respectively, where we have the commutation relations $[\hat a_u^\dagger,\hat a_v]=\delta_{uv}$. Here, $\delta_{uv}$ is the Kronecker delta. For notational simplicity, we write the annihilation and creation operators of all the $N$ modes as a vector of length $2N$ as
\begin{align}
    \vv{\hat{\zeta}} &= (\hat{a}_1,\ldots ,\hat{a}_N,\hat{a}^\dagger_1,\ldots ,\hat{a}^\dagger_N)\\
    &\equiv \left(\hat{\zeta}_1,\hat{\zeta}_2 \ldots \hat{\zeta}_N,\hat{\zeta}_{N+1},\ldots,\hat{\zeta}_{2N}\right).
\end{align}
A Gaussian state can be uniquely specified by its covariance matrix $\sigma$ and its mean vector $\vv{\beta}$. 
In our convention, the covariance matrix corresponding to a state $\rho$ is defined as 
\begin{equation}
    \sigma_{uv} = \frac{1}{2}\mathrm{Tr}\left[\rho\{\hat{\zeta}_u,\hat{\zeta}^\dagger_v\}\right] - \mathrm{Tr}\left[\rho\hat{\zeta}_u\right]\mathrm{Tr}\left[\rho\hat{\zeta}^\dagger_v\right],
\end{equation}
where $\{\cdot,\cdot \}$ represents the anti-commutator; and the mean vector is a column vector defined as
\begin{equation}
\vv{\beta} = \mathrm{Tr}[\rho\vv{\hat{\zeta}}].
\end{equation}

An electronic transition of a molecule defines a new set of vibrational modes which are displaced, distorted and rotated with respect to the vibrational modes of the ground vibronic states.
As the electronic excitation of the molecule may change forces between atoms as well as their mutual positions, the multi-mode HO corresponding to the excited state is usually expressed in a different coordinate system. If written in terms of normal coordinates, the change of the coordinate system is described by a linear relation \cite{Duschinsky}
\begin{equation}
    q'=Uq+d,
    \label{eq:Duschinsky}
\end{equation}
where $U$ is an orthogonal matrix referred to as the Duschinsky matrix, $d$ is the displacement vector, and $q$ and $q'$ are the initial and final coordinates respectively. These parameters may be determined based on {\it ab initio} calculations of molecular structures. 

In Heisenberg picture, the transition between the two states can be expressed in terms of a transformation of the ladder operators of the 
multi-mode quantum harmonic oscillator.
This transformation is a Bogoliubov transformation, and is strictly related to the transformation of the normal coordinates given in \eqref{eq:Duschinsky}. It was originally derived by Doktorov \cite{Doktorov} and used recently to show the analogy between the vibronic molecular system and Gaussian boson sampling \cite{guzik2015}. The transformation is
\begin{align}
\hat{a}^{'\dagger} &= \frac{1}{2}\left(J - (J^T)^{-1}\right)\hat{a}+ \frac{1}{2}\left(J + (J^T)^{-1}\right)\hat{a}^{\dagger}+ \frac{\vv{\delta}}{2}, \label{eqn:boug_trans}\\
&= \alpha \hat{a} + \beta \hat{a}^{\dagger} + \frac{\vv{\delta}}{2}, \label{eq:alphabetadelta}
\end{align}
with 
\begin{align}
    J &= \Omega' U \Omega^{-1},\\
     \vv{\delta} &=  \frac{\Omega' d}{\sqrt{\hbar}} ,\\
    \Omega' &= \mathrm{diag}\left(\omega_1',\ldots, \omega_N'\right)^{\frac{1}{2}}, \\
    \Omega &= \mathrm{diag}\left(\omega_1,\ldots, \omega_N\right)^{\frac{1}{2}},
\end{align}
where $U$ is the Duschinsky rotation matrix and $d$ the displacement vector from   \eqref{eq:Duschinsky}, and  $\left\{\omega_k'\right\}$ and $\left\{\omega_k\right\}$ are the harmonic angular frequencies of the final and initial states respectively. These variables are specified for a given molecule. 

At any given temperature, the vibrational modes of the ground state of the molecule are in a thermal state $\rho_i$ which is a Gaussian state. Since the evolution of the vibrational modes is specified by the Bogoliubov transformation in \eqref{eqn:boug_trans}, the final state of the molecule $\rho_f$ is also a Gaussian state. 
Therefore, we can apply the Gaussian formalism in order to find the covariance matrix and mean vector of the final Gaussian state. The evolution of the covariance matrix is given as
\begin{equation}
    \sigma_f = S\sigma_i S^T, 
\end{equation}
where $\sigma_f$ is the covariance matrix corresponding to the final state $\rho_f$, and $\sigma_i$ is the covariance matrix corresponding to the initial state $\rho_i$. In our convention,
\[S = \begin{bmatrix}
\alpha & \beta\\
\beta^* & \alpha^*
\end{bmatrix},\]
where $\alpha$ and $\beta$ are defined as in   \eqref{eq:alphabetadelta}.
Since $\alpha$ and $\beta$ are real, we can simplify the evolution of the covariance matrix as
\begin{equation} \label{eq:covariance}
    \sigma_f = S\sigma_i S.
\end{equation}

Let $\vv{d}_f$ be the mean vector corresponding to the state $\rho_f$, and let $\vv{d}_i$ be the mean vector corresponding to the state $\rho_i$. Then the mean vector evolves as
\begin{equation} \label{eq:mean}
    \vv{d}_f = S \vv{d}_i + \frac{\vv{\delta}}{\sqrt{2}},
\end{equation}
where $\vv{\delta}$ is defined in   \eqref{eq:alphabetadelta}.
In this work, we assume that the initial ground state 
is at $0$ K, and is thus a vaccuum state. 
Therefore, the initial covariance matrix $\sigma_i$ is the identity matrix, and the initial displacement $d_i$ is a null vector. We then evolve the state according to   \eqref{eqn:boug_trans} with the parameters specified by the molecule under consideration. Finally, \eqref{eq:covariance} and   \eqref{eq:mean} completely specify the final state. 

To obtain the photon number distribution of a few modes of a Gaussian state, we find the marginal states corresponding to the considered modes. These marginal states are also Gaussian. We can then obtain its covariance matrix by simply eliminating the rows and columns of the original state corresponding to the modes that are not considered. Similarly, we can also construct the mean vectors of marginal states. 

The following subsection describes the procedure of obtaining photon number distribution of these marginal states.


\subsection{Photon number distributions of Gaussian states}\label{sec:photonnumber}

A technique for calculating probabilities of photon distributions of a Gaussian state was developed in \cite{quesada2018,Kruse} and made concise in \cite{xanadu2019}. We now outline the method to calculate the probability of obtaining photon number distributions of a Gaussian state $\rho$ with covariance matrix $\sigma$ and mean vector $\vv \beta$. 
For notational simplicity, we define $\textbf{X}$ as a $2N \times 2N$ block matrix: 
\begin{equation}
    \boldsymbol{X} = \begin{bmatrix}
                        0 & \mathbb{1}_N \\
                        \mathbb{1}_N & 0
                     \end{bmatrix}. 
\end{equation}
We also define $\boldsymbol{\sigma}_Q$ from the covariance matrix $\sigma$ as:
\begin{equation}
\boldsymbol{\sigma}_Q = \sigma + \frac{1}{2}\mathbb{1}_{2N}.
\end{equation}
Following \cite{xanadu2019}, we can then define
\begin{align}
    \boldsymbol{D} & =   \boldsymbol{X}\left(\mathbb{1}_{2N} - \boldsymbol{\sigma}_Q^{-1} \right),    \label{eq:D}
 \\
    \gamma^T &= \beta^T\boldsymbol{\sigma}_Q^{-1}.
    \label{eq:gamma}
\end{align}
In order to obtain the probability of a photon number distribution $\boldsymbol{\vv{n}} = (n_1 \cdots n_N)$ from the Gaussian state characterized by $\sigma_Q$, we have the following steps:
\begin{enumerate}
\item Calculate matrix $\boldsymbol{B}$ as follows: For each $i = 1, 2, \cdots N$, the $i^{th}$ row as well as the $(N+i)^{th}$ row of $\boldsymbol{D}$ are repeated $n_i$ times; and the $i^{th}$ column as well as the $(N+i)^{th}$ column of $\boldsymbol{D}$ are repeated $n_i$ times. The resulting matrix $\boldsymbol{B}$ is a $2\Bar{n}$ dimensional square matrix where $\Bar{n} = \sum n_i$.

\item Define a vector $\Bar{\gamma}$ of length $\Bar{n}$ as follows:  For each $i = 1, 2, ... N$, the $i^{th}$ element as well as $(N + i) ^{th}$ element of  $\gamma$ is repeated $n_i$ times. 

\item Replace the diagonal entries of $\boldsymbol{B}$ with $\Bar{\gamma}$ so as to obtain a matrix $\boldsymbol{C}$ .
\end{enumerate}

The probability of obtaining a photon distribution  $\boldsymbol{\vv{n}}$  is then 
\begin{equation}
\mathrm{Prob}(\boldsymbol{\vv{n}})_{\rho}=\mathrm{F\times lhaf}(\boldsymbol{C}) \label{eq:prob}
\end{equation}
where 
\begin{equation}
    \mathrm{F} = \frac{\mathrm{exp}\left(-\frac{1}{2} \vec{\beta}^\dagger \boldsymbol{\sigma}_Q^{-1} \vec{\beta}\right)}{\sqrt{\mathrm{det}(\boldsymbol{\sigma}_Q)}\prod_{i=1}^N n_i!},
    \label{eq:T}
\end{equation}
and the function lhaf is the loop-Hafnian defined as 
\begin{align}
   \mathrm{lhaf(G)} &= \sum_{M\in W}\prod_{(i,j)\in M} G_{ij},
\end{align}
where $W$ is the set of all the perfect matchings of a graph $G$ with loops. A detailed example of this construction is shown in Appendix \ref{sec:example}. 

Calculating loop-Hafnians of large matrices is not efficient \cite{quesada2018}. In our approach, we restrict to two-mode Gaussian states and limit the number of photons to three photons per mode. This makes the calculation of loop-Hafnians tractable. The algorithm we use for calculating loop-Hafnians was obtained from \cite{loophafnian}. 

As can be observed, $\boldsymbol{C}$ has several repeated columns and rows. In such cases, one can also use a tailored formula for structured matrices as given in \cite{kan} which allows a faster calculation of loop-Hafnians.  

The method outlined above is not defined when we are interested in finding the vacuum probability of obtaining vacuum in the final Gaussian state. This is because the above procedure produces a zero-dimensional matrix $\boldsymbol{C}$. Fortunately, the vacuum probability is the overlap of two Gaussian states for which there is an efficient formula \cite{calsamiglia}. For a $N$-mode Gaussian state $\rho$ with a covariance matrix $\sigma$ and a mean vector $\vec{\beta}$, the overlap with vacuum is given as
\begin{equation}
    \mathrm{Tr}\left[\rho|\boldsymbol{0}\rangle\langle \boldsymbol{0}|\right] =\frac{\exp\left[-\frac{1}{2}\vec{\beta}^T\left(\sigma + \frac{1}{2}\mathbb{1}_{2N}\right)^{-1}\vec{\beta}\right]}{\sqrt{\mathrm{det(\sigma + \frac{1}{2}\mathbb{1}_{2N})}}} .
    \label{vacuumoverlap}
\end{equation}
This result could also have been obtained from \eqref{eq:prob} by defining loop-Hafnians of zero-dimensional matrices as one. 
\subsection{Polynomial time matching pursuit}\label{sec:ptmp}

Polynomial time matching pursuit (PTMP) is a modification of the standard matching pursuit iterative algorithm \cite{mallat} developed in the context of compressive sensing for finding a sparse high-dimensional signal $x$ that fits a given low-dimensional measurement data $y$. The transformation from $y$ to $x$ is given by a known rectangular measurement matrix $A$, i.e., $y=Ax$. In the case under consideration $x$ is the vector of FCFs, and $y$ is a vector of marginal photon number distributions. Then, $A$ is defined as in \cite{roga}. For completeness, we provide the exact form of $A$ in Appendix \ref{sec:measurement}. 
\par
PTMP is applicable when $y$ consists of   nearest neighbor marginal distributions. The algorithm is as follows:
\begin{enumerate}
    \item Initialization: Define residue $r^0=y$ and the reconstructed vector $x^0$ as the zero vector.
    \item Support detection: In step $i$, find the index $t$ of the column of $A$ with the maximum overlap with the residue $r^{i-1}$, i.e., $t = \operatorname{arg\,max}_{t'} (A^Tr^{i-1})_{t'}$. 

    \item Updating: Update the residue as $r^i=r^{i-1}-sA_{t}$, where $s$ is a chosen step size and $A_{t}$ is a column of $A$ detected in the previous step. Also, update the $t^{th}$ entry of the reconstructed vector as $x^i_t=x^{i-1}_t+s$.
\end{enumerate}

We continue the iteration until a stopping criterion is met. For us, this criterion is when the elements of $x$ sum up to one. This ensures that the sum of the transition probabilities is unity. We note that this modification of matching pursuit relies on the non-negativity of vectors and matrices involved. However, the procedure is easily adaptable to the general case.

The bottleneck of this algorithm is the support detection part. Fortunately, when we consider only marginal states of nearest neighbors, $A^Tr$ has the structure \cite{roga}
\begin{equation}
    A^Tr=\sum_{m}h_{m,m+1},
\end{equation}
where the column vector $h_{m,m+1}$ is defined as
\begin{equation}
    \left( {\bf 1}^{\otimes m-1}\otimes(r_{0_m,0_{m+1}},r_{0_m,1_{m+1}},r_{0_m,2_{m+1}}...)\otimes{\bf 1}^{\otimes N-m-1}  \right)^T
\end{equation}
where $N$ is the total number of modes and $r_{n_m,n'_{m+1}}$ denotes the probability of having $n$ photons in $m^{th}$ mode and $n'$ photons in $(m+1)^{th}$ mode. Here, ${\bf 1}^{\otimes l}$ is a vector of length $K^l$ with all entries $1$, where $K-1$ is the maximum number of photons in a mode. So, $A^Tr$ has the structure of the nearest neighbour interaction Hamiltonian for $1-D$ spin chains \cite{Ising}. 

The configuration of spins that maximizes the energy of this system, and thereby the index of the maximum entry of vector $A^Tr$, can be found efficiently using the following iterative algorithm \cite{schuch}:  
\begin{eqnarray}
E_1(i_1)&=&0 \label{isingalg}\\
E_k(i_k)&=&\max\left[E_{k-1}(i_{k-1})+H_{k-1,k}(i_{k-1},i_k)\right],\ \  k\geq 2,\nonumber
\end{eqnarray}
where matrix $H_{k-1,k}$ with two arguments $i_{k-1},i_k$ is defined by
\begin{equation}
    H_{k-1,k}(i_{k-1},i_k)=r_{(i-1)_{k-1},(i-1)_{k}}.
\end{equation}
In the original algorithm $E_k(i_k)$ is the optimized energy of spins $1,...,k$ assuming that spin $k$ is in state $i_k$. In each step, for each column ($i_k$) the maximization in \eqref{isingalg} allows us to find the index of the maximum entry in this column. This index would indicate the state of $k-1$ spins if the $k^{th}$ spin was found in state $i_k$ in the final solution. In each iteration, for each state of the currently considered spin, we have a unique sequence of states of previous spins that maximizes the energy. The final maximization determines the correct sequence of states of entire spin chain. This enables us to find the index of the maximum entry of the vector that we are concerned with. The efficiency of PTMP follows from the ability to efficiently find the maximum energy configuration of a classical spin chain in 1-D.

This procedure allows us to efficiently find the approximate distribution of sparse FCFs of molecules from the photon number distributions of nearest neighbor marginals. By sparse FCFs, we mean that either only a few FCFs are non-zero or that just a few FCFs are significantly higher  than the rest. 

\section{Results: Simulation of FC factors}\label{sec:results}
We now demonstrate the algorithm described in Section~\ref{sec:ptmp} for formic acid and thymine.
\subsection{Formic acid}\label{sec:formicacid}
As an example, let us consider formic acid. In particular, we analyze its 7-mode symmetry block with the normal mode frequencies
\begin{equation}
(3629.9;3064.9;1566.5;1399.7;1215.3;1190.9;496.3)[\mathrm{cm^{-1}}]
\label{fa_freq}
\end{equation}
for the electronic transition $(1^1A'')\rightarrow (1^2A')$. The parameters defining this transition are given in \cite{guzik2015}.

We  obtain the marginal probabilities of photon numbers in adjacent modes, considering up to three photons per mode. This result was then used to reconstruct the distribution of FCFs using PTMP. The step size in our algorithm is chosen as $s=0.01$, which necessitates no more than $k=100$ iterations of our algorithm. We do not need to change the parameters $s$ and $k$ for larger molecules.

\begin{figure}
    \includegraphics[scale=0.4]{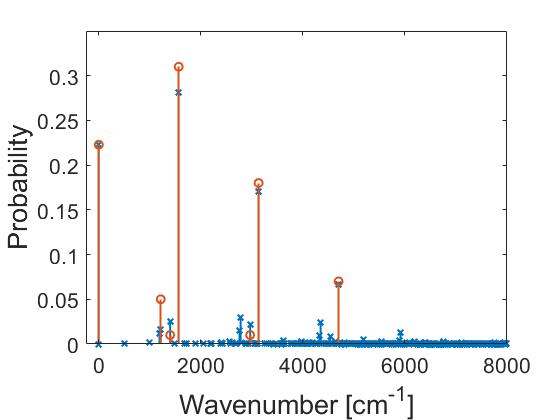}
    \caption{Franck-Condon factors for transitions between the electronic ground state $(1^1A'')$ and vibrational states in the electronic excited state $(1^2A')$ of formic acid at 0 K. The blue lines with crosses is the spectrum obtained by the exact Doktorov recursive method, the orange lines with circles is the spectrum reconstructed from the nearest neighbors marginal distributions by PTMP. The reconstruction quality ( the $l1$-norm between the reconstructed distribution and the exact one) is 0.2991.}
    \label{fig:fa_energy}
\end{figure}

As we consider a small molecule and limit the maximum number of photons per mode, the exact approach by Doktorov is computationally tractable using a standard desktop computer. This enables us to  plot the result of the reconstruction by PTMP together with the exact spectrum in Figure \ref{fig:fa_energy}. As a reconstruction quality measure, we use the $l1-$norm between the exact FCFs ($p$) and the reconstructed FCFs ($q$) defined as  $D_{tr}=\sum_i|p_i-q_i|$.

The main lines shown in Figure \ref{fig:fa_energy} are described in Table \ref{tab:fa}. We observe that the main lines correspond to one, two, and three photons excited in a particular mode. Our reconstruction also allows us to recognize lines corresponding to  simultaneous excitation in two different modes. In our reconstruction of FCFs,  we could have restricted the total number of excitations to just a few or restricted the excitations to only  a few modes. However, the method we use does not use any such assumption.

\begin{table}[]
\begin{tabular}{l|c|l}
 Wavenumber $[\mathrm{cm^{-1}}]$ &  Photon number state & Probability   \\ \hline
0 & 0000000 & 0.2228  \\
1566.5 & 0010000 & 0.31 \\
3132.9 & 0020000 & 0.18 \\
4699.4 & 0030000 & 0.07 \\ 
1215.3 & 0000100 & 0.05 \\
1399.7 & 0001000 & 0.01 \\
2966.1 & 0011000 & 0.01
\end{tabular}
\caption{Main FCFs of formic acid reconstructed from marginal distributions of photon numbers in the nearest neighbor modes by PTMP. The modes correspond to the symmetry block with seven normal modes with wave numbers in (\ref{fa_freq}). }
\label{tab:fa}
\end{table}    

\begin{figure}
    \includegraphics[scale=0.4]{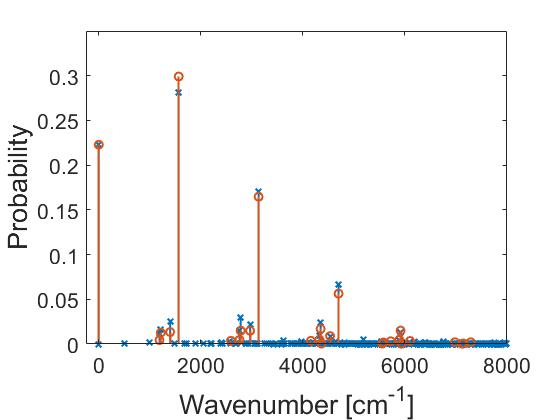}
    \caption{Franck-Condon factors for transitions between the electronic ground state $(1^1A'')$ and vibrational states in the electronic excited state $(1^2A')$ of formic acid at 0 K. The blue lines with crosses is the spectrum obtained by the exact Doktorov recursive method, the orange lines with circles is the spectrum of minimum $l1$-norm that fits the nearest neighbors marginal distributions. The reconstruction quality (the $l1$-norm between the reconstructed distribution and the exact one) is 0.1988.}
    \label{fig:fa_energy_cvx}
\end{figure}

Since formic acid is small enough that its FCFs can be reconstructed by other methods, we compare the performance of PTMP with another compressive sensing method. This method finds a solution which fits the marginal distributions while  minimizing it's $l1$-norm. For this, we use the constrained $l1$-norm minimization in MATLAB using the CVX package \cite{cvx}. The spectrum thus obtained is plotted in Figure~\ref{fig:fa_energy_cvx}.

From Figure \ref{fig:fa_energy} and Figure \ref{fig:fa_energy_cvx}, we observe that the quality of the reconstruction is better while using norm-minimization (0.1988) than while using PTMP (0.2911). It is known that norm-minimization  provides better solutions than first order iterative algorithms like matching pursuit. However,  norm-minimization  requires more memory than first order iterative algorithms. Further,  norm-minimization scales badly with the size of the problem. In particular, it is inefficient for finding the distribution of FCFs of thymine which we consider next. In contrast, PTMP does not experience computational time or memory problems. 

\subsection{Thymine}\label{sec:thymine}
As the second example we consider thymine. Its vibrational degrees of freedom can be decoupled into two separate blocks of $13$ normal modes and $26$ normal modes. We consider the transitions at $0$ K between the ground electronic state $(1^1A'')$ and $26$ vibrational modes of the excited state $(1^2A')$ with normal frequencies as follows:
\begin{eqnarray}
&&(3535.0;3511.8;3195.7;3150.6;2996.8;1833.9;1739.7;\nonumber\\ 
&&1575.5;1531.1;1474.4;1442.6;1380.0;1353.8;1315.0;\nonumber\\
&&1271.3;1216.1;1187.3;995.9;893.5;766.6;690.7;581.6;\nonumber\\
&&532.2;444.4;392.2;293.8)\mathrm{[cm^{-1}]}.
\label{thymine_freq}
\end{eqnarray}
Obtaining the spectrum of thymine is more challenging with standard techniques. This is because even if we restrict ourselves to transitions with no more than three photons per mode, then the number of all possible FCFs is $4^{26}$. Due to this size, dealing with this  molecule using standard methods such as the exact recursive Doktorov method is infeasible on standard desktop computers. Also, $l1$-norm minimizer can not deal with problems of this size.   

\begin{figure}
    \includegraphics[scale=0.25]{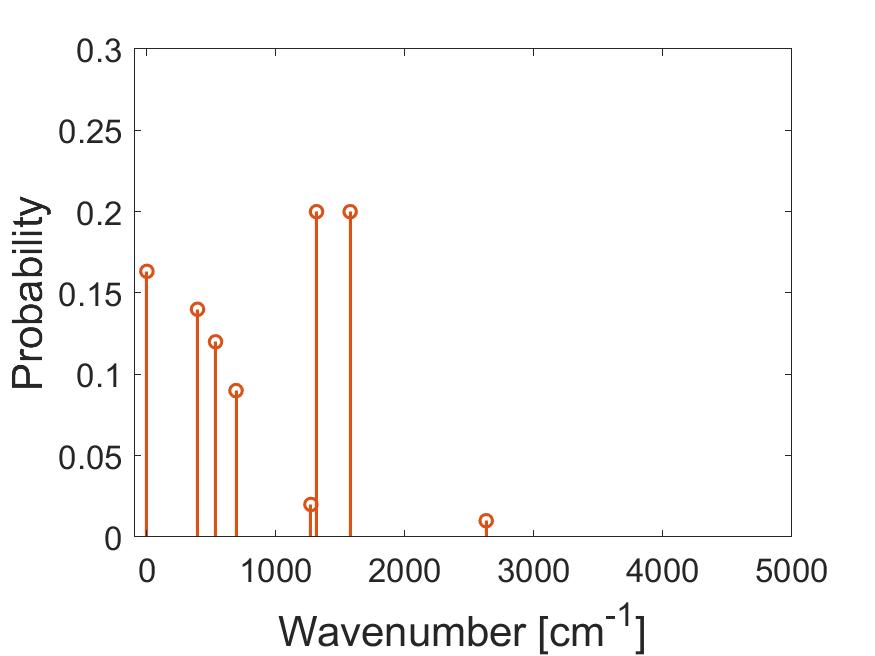}
    \caption{Franck-Condon factors for transitions between the electronic ground state $(1^1A'')$ in $0$ K and $26$-vibrational modes of an excited electronic state $(1^2A')$ of thymine. This spectrum is recovered by PTMP from  marginal distributions for all nearest neighbor modes.}
    \label{fig:thymine}
\end{figure}

Based on the parameters of the transition provided in \cite{Huh,Jankowiak}, we obtain the Gaussian state associated with the excited state of the transition. Next, we find the marginal states for nearest neighbor modes, and then find their photon number distributions up to three photons per mode. Finally, we apply the PTMP algorithm to find the FCFs. We show the spectrum of FCFs of thymine in Figure~\ref{fig:thymine} and in Table~\ref{tab:th}. 

The approximate spectrum of FCFs of thymine produced by PTMP agrees well with the spectra produced by other methods \cite{Jankowiak}. Although our method is not as precise as the method given in \cite{Jankowiak}, the computation is faster. In particular, the calculations used to produce the spectrum given in Figure~\ref{fig:thymine} take only a few seconds on a standard desktop computer. Observing the spectra of FCFs of thymine, in contrast to that of formic acid, we observe that the main lines correspond to single photon transitions for different modes.  

\begin{table}[]
\begin{center}
\begin{tabular}{l|c|l}
 Wavenumber $[\mathrm{cm^{-1}}]$ &  Photon Number State & Probability   \\ \hline
0 & 0 photon in all modes & 0.1633  \\
392.2 & 1 photon in $25^{\mathrm{th}}$ mode & 0.14 \\
532.2 & 1 photon in $23^{\mathrm{rd}}$ mode & 0.12 \\
690.7 & 1 photon in $21^{\mathrm{st}}$ mode & 0.09 \\ 
1315.0 & 1 photon in $14^{\mathrm{th}}$ mode & 0.02 \\
1575.5 & 1 photon in $8^{\mathrm{th}}$ mode & 0.02
\end{tabular}
\caption{Main FCFs of thymine reconstructed from marginal distributions of photon numbers in the nearest neighbor modes by PTMP. The modes correspond to the symmetry block with 26 normal modes with wave numbers in (\ref{thymine_freq}). }
\label{tab:th}
\end{center}
\end{table}

Let us comment on the sparsity assumption for our method. We can observe that the exact spectrum of FCFs of thymine is not sparse \cite{Jankowiak,guzik2015}. However, there are six lines with significantly higher probabilities than the other lines of the spectrum. This situation is similar to a sparse signal in the presence of noise. We observe that even though many small probability contributions are present, we are still able to reconstruct the most significant lines of the spectrum. This shows that our algorithm does not demand strict sparsity of FCFs. However, recognizing the lines with lower probabilities is out of the reach and beyond the scope of this approach. 

\section{Conclusions}

In this paper, we develop an efficient classical method to approximately reconstruct the distribution of FCFs of large molecules. This reconstruction is possible with an efficient compressive sensing algorithm which uses marginal photon number probability distributions as compressively sensed data. 
We define and test the performance of the polynomial time matching pursuit (PTMP) algorithm which allows us to efficiently reconstruct the main peaks of molecular vibronic spectra even for large molecules.

Our method is restricted to spectra which are sparse or spectra with just a few FCFs significantly larger than the rest. In the case of decreasing sparsity we would need more marginal distributions to reliably reconstruct significant FCFs. In such a case, instead of two nearest neighbor modes we could consider three or more nearest neighbor modes.  PTMP can be easily adapted to and will still be efficient in this case. However, the computational time for calculating loop-Hafnians increases exponentially with the number of modes which restricts the method. 

Since we study FCFs at temperature $0$ K, we can assume that each mode of the final state is occupied by only few photons. However when the temperature is finite, the initial state of the molecule is a thermal state with the average number of photons in a mode obeying the Bose-Einstein distribution. For a typical molecule (say formic acid), even for a relatively low temperature of about $1$ K, the average number of photons in a mode is of the order of $10^3$. This implies that marginal distributions containing relatively few photons are improbable. In such cases, although the matrices of whose loop-Hafnians are to be calculated are huge, they are highly structured matrices with only a few dissimilar rows/columns. Then, we can employ the method of Kan \cite{kan} to calculate such loop-Hafnians. This can be a subject of future studies.

Finally, the method of FCFs recovered from marginal distributions by compressive sensing may be improved by introducing algorithms other than PTMP. This can include gradient pursuit introduced in \cite{gradientpursuit} as discussed in \cite{roga}. We also think that orthogonal matching pursuit \cite{zhang} can be adapted for this purpose.\\

Acknowledgements: The authors thank Joonsuk Huh for sharing the Duschinsky matrix for thymine. The authors also thank Takafumi Ono and Jonathan P. Dowling for helpful discussions.

KVJ acknowledges the support from NSF and from Louisiana State University System Board of Regents via an Economic Development Assistantship. EK acknowledges support from US Office of Naval Research. WR and MT acknowledge the support of JST CREST Grant No. JPMJCR1772.

\bibliographystyle{unsrt}
\bibliography{FC-factors}

\begin{thebibliography}{10}

\bibitem{hachmann}
J.~Hachmann, R.~Olivares-Amaya, S.~Atahan-Evrenk, C.~Amador-Bedolla, R.~S.
  S\'{a}nchez-Carrera, A.~Gold-Parker, L.~Vogt, A.~M. Brockway, and
  A.~Aspru-Guzik.
\newblock The {H}arvard clean energy project: large-scale computational
  screening and design of organic photovoltaniaics on the world community grid.
\newblock {\em J. Phys. Chem. Lett.}, 2:2241--2251, August 2011.

\bibitem{butler}
H.~J. Butler, B.~Bird L.~Ashton, G.~Cinque, K.~Curtis, J.~Dorney,
  K.~Esmonde-White, N.~J. Fullwood, B.~Gardner, P.~L. Martin-Hirsch, M.~J.
  Walsh, M.~R. McAinsh, N.~Stone, and F.~L. Martin.
\newblock Using {R}aman spectroscopy to characterize biological materials.
\newblock {\em Nature Protocols}, 11:664--687, March 2016.

\bibitem{gross2000}
M.~Gross, D.~C. M\"uller, H.-G. Nothofer, U.~Scherf, D.~Neher, C.~Br\"auchle,
  and K.~Meerholz.
\newblock Improving the performance of doped $\pi$-configurated polymers for
  use in organic light-emitting diodes.
\newblock {\em Nature}, 405:661--665, June 2000.

\bibitem{philips2019}
D.~S. Phillips, M.~Walschaers, J.~J. Renema, I.~A. Walmsley, N.~Treps, and
  J.~Sperling.
\newblock Benchmarking of {G}aussian boson sampling using two-point
  correlators.
\newblock {\em Phys. Rev. A}, 99:023836, Feb 2019.

\bibitem{Doktorov}
E.V.Doktorov, I.A.Malkin, and V.I.Ma\'{n}ko.
\newblock The importance of the electron spectrum in multi atomic molecules.
  concerning {F}ranck-{C}ondon principle.
\newblock {\em Journal of Molecular Spectroscopy}, 64:302--326, February 1977.

\bibitem{Rabidoux}
S.~M. Rabidoux, V.~Eijkhout, and J.~F. Stanton.
\newblock A highly-efficient implementation of the {D}oktorov recurrence
  equations for {F}ranck-{C}ondon calculations.
\newblock {\em J. Chem. Theory Comput.}, 12:728--739, January 2016.

\bibitem{Jankowiak}
J.~L.~Stuber H.-C.~Jankowiak and R.~Berger.
\newblock Vibronic transitions in large molecular systems: Rigorous
  prescreening conditions for {F}ranck-{C}ondon factors.
\newblock {\em The Journal of Chemical Physics}, 127:234101, December 2007.

\bibitem{santoro}
Fabrizio Santoro, Roberto Improta, Alessandro Lami, Julien Bloino, and Vincenzo
  Barone.
\newblock Effective method to compute franck-condon integrals for optical
  spectra of large molecules in solution.
\newblock {\em The Journal of Chemical Physics}, 126(8):084509, 2007.

\bibitem{Ruhoff}
{Peder Thusgaard} Ruhoff and {Mark A} Ratner.
\newblock Algorithms for computing franck-condon overlap integrals.
\newblock {\em International Journal of Quantum Chemistry}, 77(1):383--392, 3
  2000.

\bibitem{quesada2018}
N.~Quesada.
\newblock Franck-{C}ondon factors by counting perfect matchings of graphs with
  loops.
\newblock {\em The Journal of Chemical Physics}, 150:164113, April 2019.

\bibitem{Candes2}
E.~J. {Candes} and T.~{Tao}.
\newblock Decoding by linear programming.
\newblock {\em IEEE Transactions on Information Theory}, 51(12):4203--4215,
  December 2005.

\bibitem{Donoho}
D.~L. Donoho.
\newblock {\em Communications on Pure and Applied Mathematics}, 59(7):907--934,
  2006.

\bibitem{Candes1}
E.~J. {Candes} and T.~{Tao}.
\newblock Near-optimal signal recovery from random projections: Universal
  encoding strategies?
\newblock {\em IEEE Transactions on Information Theory}, 52(12):5406--5425,
  December 2006.

\bibitem{Baraniuk}
Richard Baraniuk, Mark Davenport, Ronald DeVore, and Michael Wakin.
\newblock A simple proof of the restricted isometry property for random
  matrices.
\newblock {\em Constr Approx}, 28:253--263, December 2008.

\bibitem{Candes3}
E.~J. {Candes}, J.~{Romberg}, and T.~{Tao}.
\newblock Robust uncertainty principles: exact signal reconstruction from
  highly incomplete frequency information.
\newblock {\em IEEE Transactions on Information Theory}, 52(2):489--509,
  February 2006.

\bibitem{mallat}
S.~G. Mallat and Z.~Zhang.
\newblock Matching pursuits with time-frequency dictionaries.
\newblock {\em IEEE Transactions on Signal Processing}, 41:3397, October 1993.

\bibitem{davis}
G.~Davis, S.~Mallat, and Avellaneda M.
\newblock Adaptive greedy approximations.
\newblock {\em Constructive approximation}, 13:57–98, 1997.

\bibitem{gradientpursuit}
T.~{Blumensath} and M.~E. {Davies}.
\newblock Gradient pursuits.
\newblock {\em IEEE Transactions on Signal Processing}, 56(6):2370--2382, June
  2008.

\bibitem{blumensath2}
T.~Blumensath and M.~E. Davies.
\newblock Iterative thresholding for sparse approximations.
\newblock {\em Journal of Fourier Analysis and Applications}, 14:629--654,
  2008.

\bibitem{Draganic}
A.~Draganic, I.~Orovic, and S.~Stankovic.
\newblock On some common compressive sensing algorithms and applications -
  review paper.
\newblock {\em Facta Universitatis, Series: Electronics and Energetics},
  30(4):477--510, 2017.

\bibitem{zhang}
T.~Zhang.
\newblock Sparse recovery with orthogonal matching pursuit under rip.
\newblock {\em IEEE Trans. on Information Theory}, 57:6215--6221, 2011.

\bibitem{romberg}
J.~Romberg.
\newblock Imaging via compressive sampling.
\newblock {\em IEEE Signal Processing Magazine}, 25:14--20, 2008.

\bibitem{mihajlovic}
R.~Mihajlovic, M.~Scekic, A.~Draganic, and S.~Stankovic.
\newblock An analysis of cs algorithms efficiency for sparse communication
  signals reconstruction.
\newblock {\em 3rd Mediterranean Conference on Embedded Computing, MECO}.

\bibitem{lustig}
M.~Lustig, D.~Donoho, and J.~M. Pauly.
\newblock Sparse mri: The application of compressed sensing for rapid mr
  imaging.
\newblock {\em Magnetic Resonance in Medicine}, 58:1182--1195, 2007.

\bibitem{gross}
David Gross, Yi-Kai Liu, Steven~T. Flammia, Stephen Becker, and Jens Eisert.
\newblock Quantum state tomography via compressed sensing.
\newblock {\em Phys. Rev. Lett.}, 105:150401, October 2010.

\bibitem{guzik2015}
Joonsuk Huh, Gian~Giacomo Guerreschi, Borja Peropadre, Jarrod~R. McClean, and
  Al\'an Aspuru-Guzik.
\newblock Boson sampling for molecular vibronic spectra.
\newblock {\em Nature Photonics}, 9:615, August 2015.

\bibitem{Hamilton2017}
Craig~S. Hamilton, Regina Kruse, Linda Sansoni, Sonja Barkhofen, Christine
  Silberhorn, and Igor Jex.
\newblock Gaussian boson sampling.
\newblock {\em Phys. Rev. Lett.}, 119:170501, Oct 2017.

\bibitem{weedbrook}
Christian Weedbrook, Stefano Pirandola, Ra\'ul Garc\'{\i}a-Patr\'on, Nicolas~J.
  Cerf, Timothy~C. Ralph, Jeffrey~H. Shapiro, and Seth Lloyd.
\newblock Gaussian quantum information.
\newblock {\em Rev. Mod. Phys.}, 84:621--669, May 2012.

\bibitem{Kruse}
R.~Kruse, C.~S. Hamilton, L.~Sansoni, S.~Barkhofen, C.~Silberhorn, and I.~Jex.
\newblock A detailed study of {G}aussian boson sampling.
\newblock arXiv:1801.07488.

\bibitem{roga}
W.~Roga and M.~Takeoka.
\newblock Classical simulation of boson sampling with sparse output.
\newblock arXiv:1904.05494.

\bibitem{Braunstein}
Samuel~L. Braunstein and Peter van Loock.
\newblock Quantum information with continuous variables.
\newblock {\em Rev. Mod. Phys.}, 77:513--577, Jun 2005.

\bibitem{Eisert}
J.~Eisert and M.~B. Plenio.
\newblock Introduction to the basics of entanglement theory in
  continuous-variable systems.
\newblock {\em International Journal of Quantum Information}, 01(04):479--506,
  2003.

\bibitem{Braunstein2}
Samuel~L. Braunstein and Peter van Loock.
\newblock Quantum information with continuous variables.
\newblock {\em Rev. Mod. Phys.}, 77:513--577, Jun 2005.

\bibitem{Ferraro}
A.~Ferraro, S.~Olivares, and {Matteo G. A.} Paris.
\newblock {\em Gaussian States in Quantum Information}.
\newblock Napoli Series on physics and Astrophysics. 2005.

\bibitem{Duschinsky}
F.~Duschinsky.
\newblock The importance of the electron spectrum in multi atomic molecules.
  {C}oncerning {F}ranck-{C}ondon principle.
\newblock {\em Acta Physicochim. URSS}, 7, 1937.

\bibitem{xanadu2019}
N.~Quesada, L.~G. Helt, J.~Izaac, J.~M. Arrazola, R.~Shahrokhshahi, C.~R.
  Myers, and K.~K. Sabapathy.
\newblock Simulating realistic non-{G}aussian state preparation.
\newblock arXiv:1905.07011.

\bibitem{loophafnian}
Andreas Bj\"{o}rklund, Brajesh Gupt, and Nicol\'{a}s Quesada.
\newblock A faster hafnian formula for complex matrices and its benchmarking on
  a supercomputer.
\newblock {\em J. Exp. Algorithmics}, 24(1):1.11:1--1.11:17, June 2019.

\bibitem{kan}
Raymond Kan.
\newblock From moments of sum to moments of product.
\newblock {\em Journal of Multivariate Analysis}, 99(3):542 -- 554, 2008.

\bibitem{calsamiglia}
J.~Calsamiglia, R.~Mu\~noz Tapia, Ll. Masanes, A.~Acin, and E.~Bagan.
\newblock Quantum chernoff bound as a measure of distinguishability between
  density matrices: Application to qubit and gaussian states.
\newblock {\em Phys. Rev. A}, 77:032311, Mar 2008.

\bibitem{Ising}
E.~{Ising}.
\newblock {Beitrag zur Theorie des Ferromagnetismus}.
\newblock {\em Zeitschrift fur Physik}, 31:253--258, February 1925.

\bibitem{schuch}
Norbert Schuch and J.~Ignacio Cirac.
\newblock Matrix product state and mean-field solutions for one-dimensional
  systems can be found efficiently.
\newblock {\em Phys. Rev. A}, 82:012314, July 2010.

\bibitem{cvx}
Michael Grant and Stephen Boyd.
\newblock Cvx: Matlab software for disciplined convex programming, version 2.0
  beta.

\bibitem{Huh}
J.~Huh and M-H Yung.
\newblock Vibronic boson sampling: Generalized {G}aussian boson sampling for
  {M}olecular {V}ibronic spectra at {F}inite temperature.
\newblock {\em Scientific Reports}, 7:7462, August 2017.

\end{thebibliography}

\begin{appendix}
\section*{Appendix}

\section{Photon number distributions of a Gaussian state}\label{sec:example}

As an example of calculating the photon number distribution of a multimode Gaussian state, consider a two mode Gaussian state described by a covariance matrix $\sigma$ and a mean vector $\beta$ as follows:
\begin{align}
    \sigma &= \begin{bmatrix}
                        \sigma_{11} & \sigma_{12} & \sigma_{13} & \sigma_{14} \\
                        \sigma_{21} & \sigma_{22} & \sigma_{23} & \sigma_{24} \\
                        \sigma_{31} & \sigma_{32} & \sigma_{33} & \sigma_{34} \\
                        \sigma_{41} & \sigma_{42} & \sigma_{43} & \sigma_{44}
                        \end{bmatrix},   \\
    \beta  &= \begin{bmatrix}
                        \beta_1 &
                        \beta_2 &
                        \beta_3 &
                        \beta_4
                        \end{bmatrix}^T.
\end{align}
We are interested in the probability of obtaining two photons in mode 1 and one photon in mode 2. We first calculate the matrix $\boldsymbol{D}$ as in \eqref{eq:D}. Then we construct the 6 $\times$ 6 matrix $\boldsymbol{B}$ as
\begin{equation}
    \boldsymbol{B} = \begin{bmatrix}
                    D_{11} & D_{11} &D_{12} & D_{13} & D_{13} &D_{14} \\
                    D_{11} &D_{11} &D_{12} &D_{13} &D_{13} &D_{14} \\
                    D_{21} & D_{21} &D_{22} &D_{23} &D_{23} &D_{24} \\
                    D_{31} & D_{31} &D_{32} &D_{33} &D_{33} &D_{34} \\
                   D_{31} &D_{31} &D_{32} &D_{33} &D_{33} &D_{34} \\
                    D_{41} &D_{41} &D_{42} &D_{43} &D_{43} &D_{44} \\
                    \end{bmatrix}.
\end{equation}
Constructing $\vv \gamma$ as per   \eqref{eq:gamma}, we obtain
\begin{equation}
   \Bar{\gamma} = \begin{bmatrix}
                        \gamma_1 &
                        \gamma_1 &
                        \gamma_2 &
                        \gamma_3 &
                        \gamma_3 &
                        \gamma_4
                        \end{bmatrix}^T.
\end{equation}

Next, we replace the diagonal entries of $\boldsymbol{B}$ with $\Bar{\gamma}$  so as to obtain
\begin{equation}
    \boldsymbol{C} = \begin{bmatrix}
                         \gamma_1 & D_{11} &D_{12} & D_{13} & D_{13} &D_{14} \\
                    D_{11} &\gamma_1 &D_{12} &D_{13} &D_{13} &D_{14} \\
                    D_{21} & D_{21} &\gamma_2 &D_{23} &D_{23} &D_{24} \\
                    D_{31} & D_{31} &D_{32} &\gamma_3&D_{33} &D_{34} \\
                   D_{31} &D_{31} &D_{32} &D_{33} &\gamma_3 &D_{34} \\
                    D_{41} &D_{41} &D_{42} &D_{43} &D_{43} &\gamma_4 \\
                    \end{bmatrix}.
\end{equation}
Finally, the loop-Hafnian of the above matrix is calculated, and then used to calculate the concerned probability using   ~\eqref{eq:prob}.

\begin{section}{Measurement matrix}\label{sec:measurement}
Since our measurement matrix is large, it is imperative that we develop a particular representation of the measurement matrix such that its elements can be efficiently generated by knowing it's indices alone. This makes our algorithm efficient with respect to its memory requirement.

In order to define the measurement matrix, we first develop a representation for the vector ($x$) with all FCFs. To this end, we define vectors specified by sets of photon numbers in each mode $\{n_1,n_2,...\}$, where $n_i$ denotes the number of photons in the $i$-th mode as follows:
\begin{equation}
\begin{array}{c c c c c c}
|n_1,n_2,...)^T=&(\cdot\cdot\cdot 1 \cdot \cdot)&\otimes&(\cdot\cdot 1 \cdot\cdot\cdot)&\otimes&...\\
\ &\uparrow&\ &\uparrow& \ \\
\ &n_1+1&\ &n_2+1 &\ &...
\end{array}.
\label{tensorrepres}
\end{equation}
Here the lower indices indicate the positions of $1$ in each component of the tensor product, and the remaining entries are zeros. Using this representation, we can then decompose the vector $x$ of all FCFs of an $N$ mode system as follows:
\begin{equation}
x=\sum_{n_1,...,n_N}\alpha_{n_1,...,n_M}|n_1,...,n_N).
\end{equation}
Here, the coefficients $\alpha_{n_1,...,n_N}$ are non-negative and sum up to one.

We will now see how to define the measurement matrix in this representation. For this, let us first define the following auxiliary vectors
\begin{equation}
\begin{array}{c c c c c c}
\gamma_{n_i}=&{\bf 1}^{\otimes i-1}&\otimes&(\cdot\cdot 1 \cdot\cdot \cdot)&\otimes&{\bf 1}^{\otimes N-i}\\
\ &\ &\ &\uparrow& \ & \\
\ &\ &\ &n_i+1 &\ &
\end{array}.
\label{gammam0}
\end{equation}
where ${\bf 1}=(1,1,1,...)$ and $(\cdot\cdot 1\cdot\cdot \cdot)$ are $K$ dimensional vectors and $K-1$ is the maximum photon number in a mode. $\gamma_{n_i}$ is defined such that $\gamma_{n_i}x$ gives the marginal probability of having $n_i$ photons in the $i$-th mode. By choosing different $n_i$, we can then obtain the marginal probability distribution of the photon numbers in mode $i$. 
 
 In order to find the marginal distributions of two modes, we use the entry-wise products $\gamma_{n_i}\odot \gamma_{n_j}$ such that $(\gamma_{n_i}\odot \gamma_{n_j})x$ is just the marginal probability of simultaneously finding  $n_i$ photons in the $i^{th}$ mode and $n_j$ photons in the $j^{th}$ mode. These quantities allow us to write the marginal distributions of photonic occupations in pairs of modes. 
 
 In the problem that we are concerned, the measurement matrix $A$ consists of rows indexed by $(n_i,n_j)$  given as
 \begin{equation}
     A_{n_i,n_j} = \gamma_{n_i}\odot \gamma_{n_j}.
 \end{equation}
 This matrix describes the transition from the joint distribution of FCFs for all modes and marginal distributions for pairs of modes. Using this form of the measurement matrix, the vector of joint marginal distributions ($y$)  is given as 
\begin{equation}
y_{n_i, n_j} = A_{n_i,n_j}x.
\label{ope}
\end{equation}
We can now apply compressive sensing methods to our problem.
\end{section}

\end{appendix}

\end{document}